\pdfoutput=1
\documentclass{IEEEoj}

\usepackage{cite}
\usepackage{amsmath,amssymb,amsthm}
\usepackage{bm}
\usepackage{graphicx}
\usepackage{float}
\usepackage{array}
\usepackage{booktabs}
\usepackage[T1]{fontenc}
\usepackage[utf8]{inputenc}
\usepackage{textcomp}
\usepackage[T1]{fontenc}
\usepackage{anyfontsize} 
\AtBeginDocument{\definecolor{ojcolor}{cmyk}{0.93,0.59,0.15,0.02}}
\def\OJlogo{\vspace{-4pt}\hskip-4pt\includegraphics[height=18pt]{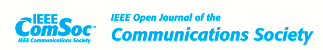}}

\newcommand{\lb}{\left}
\newcommand{\rb}{\right}

\DeclareMathOperator{\diag}{diag}
\DeclareMathOperator{\sgn}{sign}
\allowdisplaybreaks

\begin{document}

\receiveddate{XX Month, XXXX}
\reviseddate{XX Month, XXXX}
\accepteddate{XX Month, XXXX}
\publisheddate{XX Month, XXXX}
\currentdate{XX Month, 2026}
\doiinfo{OJCOMS.2026.0000000}

\title{Beamforming and RIS-Aided Ambient Backscatter Communications with {Residual-Feature} SVM Detection}

\author{AOBAKWE MAKGARAPA\IEEEauthorrefmark{1} \IEEEmembership{(Student Member, IEEE)}, BURHAN WAFAI\IEEEauthorrefmark{2} \IEEEmembership{(Graduate Student Member, IEEE)}, SANGARAPILLAI LAMBOTHARAN\IEEEauthorrefmark{3} \IEEEmembership{(Senior Member, IEEE)}, AND MAHSA DERAKHSHANI\IEEEauthorrefmark{1} \IEEEmembership{(Senior Member, IEEE)}}
\affil{Wolfson School of Mechanical, Electrical and Manufacturing Engineering, Loughborough University, Loughborough, U.K.}
\affil{East Midlands Institute of Technology, Loughborough College Group, U.K.}
\affil{Institute for Digital Technologies, Loughborough University, London, U.K.}
\corresp{CORRESPONDING AUTHOR: Aobakwe Makgarapa (e-mail: a.makgarapa@lboro.ac.uk).}
\markboth{Beamforming and RIS-Aided Ambient Backscatter Communications with Residual-Feature SVM Detection}{Makgarapa \textit{et al.}}

\begin{abstract}
Ambient backscatter communication (AmBC) enables battery-free connectivity by modulating data onto existing radio-frequency (RF) signals, eliminating the need for dedicated power sources. However, its reliability degrades when direct wireless channels are obstructed or severely faded. Reconfigurable intelligent surfaces (RISs) offer a solution by creating a favorable propagation environment through the control of the phases of incident signals, thereby strengthening wireless links.
This paper investigates an RIS-aided AmBC system that jointly exploits physical-layer reconfiguration and statistical learning to restore detection reliability under such conditions. The RIS phase profile is aligned for the source-RIS-tag link, while a multi-antenna reader applies receive beamforming to steer toward the RF source and the tag separately. At the reader, a hypothesis-based minimum mean square error (MMSE) equalizer reconstructs the ambient symbol under each candidate tag state and produces a pair of residual features, which are classified by a support vector machine (SVM) with a Gaussian kernel. We make the physical-layer-to-learning coupling explicit, i.e., RIS phase alignment and beamforming improve the signal-to-interference-plus-noise ratio (SINR), thereby rendering the residual features more separable and reducing classification error. 
Simulation results show that the proposed RIS-beamforming-SVM detector achieves substantial bit-error-rate (BER) gains over RIS-energy, SVM, and SVM-beamforming baselines across a wide SINR range, that the spectral-efficiency gains are governed more strongly by the number of RIS elements than by the number of reader antennas, and that performance saturates with a moderate RIS size, allowing near-optimal operation at reduced hardware cost.
\end{abstract}

\begin{IEEEkeywords}
Ambient backscatter, reconfigurable intelligent surface (RIS), support vector machine (SVM), beamforming, MMSE equalization, obstructed propagation, Internet of Things (IoT).
\end{IEEEkeywords}

\maketitle

\section{Introduction}
\IEEEPARstart{A}{mbient} backscatter communication (AmBC) has emerged as a promising paradigm for enabling low-power and battery-free communication in large-scale Internet of Things (IoT) networks \cite{11072257,8368232,long2020symbiotic}. Rather than generating an active carrier, an AmBC tag modulates information directly onto existing radio-frequency (RF) signals such as television, cellular, or Wi-Fi transmissions, by varying its antenna load impedance and thereby altering its reflection coefficient \cite{liu2013ambient}. Because the tag avoids power-hungry RF chains, AmBC substantially reduces energy consumption and hardware complexity, which makes it attractive for dense sensing deployments where battery replacement and maintenance are impractical \cite{10915625,8368232}.

These benefits, however, are accompanied by stringent reliability limitations that distinguish AmBC from conventional active links. The backscattered component is typically orders of magnitude weaker than the direct ambient path. The strong direct-link signal acts as self-interference that can overwhelm the tag's reflection, and the propagation environment is governed by uncontrolled ambient sources that the system designer cannot regulate \cite{10947298,zheng2020ambient,Wang2016AmbientBackscatter}. These difficulties are exacerbated in obstructed and deep-fading propagation, where the direct source-to-reader and source-to-tag links are shadowed, blocked, or deep-faded. In such regimes, the tag receives a weak signal, and the reader observes a backscatter contribution that is barely distinguishable from noise, rendering conventional energy detectors and even coherent detectors ineffective. This scenario is precisely the one in which AmBC is most needed, low-power sensors in cluttered indoor, industrial, or body-area environments, and yet is the one in which receiver processing alone is least able to help.

Reconfigurable intelligent surface (RIS) offers a complementary degree of freedom by reshaping the wireless environment in favor of the backscatter link \cite{10409611,wafai_tvt,10200237}. An RIS is a planar array of low-cost, nearly passive reflecting elements whose phases can be independently configured to combine multipath components constructively, thereby improving signal strength and extending coverage \cite{wu2019intelligent,DiRenzo2020RIS,wu2021tutorial}. Positioned between the RF source, the tag, and the reader, an RIS can create a controllable cascaded source-RIS-tag path that strengthens the signal reaching the tag and reinforces the desired component at the reader. Crucially, in the obstructed-propagation setting addressed here, when the direct source-to-tag link is weak, the RIS-assisted path can become the dominant channel link, enabling the system to recover useful signal energy that would otherwise be lost. While RIS integration is now extensively studied in active communication systems, its deployment in AmBC remains comparatively nascent, with open questions surrounding placement, phase configuration, and, most relevant here, integration with advanced detection algorithms.

In parallel, machine learning (ML) has gained increasing attention in wireless communications for tasks such as channel estimation, signal detection, and resource allocation \cite{8741198,wafai_ojcoms,ye2018power}. In AmBC, the highly dynamic interference environment and the nonlinear mixing of the direct and backscatter components challenge conventional detectors based on energy thresholds or maximum-likelihood estimation \cite{Wang2016AmbientBackscatter,zheng2020ambient}. The minimum mean square error (MMSE) detector remains a powerful and well-established baseline, but its performance can be complemented by ML approaches that exploit statistical structure in the received signal without relying entirely on accurate channel state information (CSI). Among ML classifiers, support vector machines (SVMs) are particularly well-suited to the AmBC tag-detection problem, as the decision is intrinsically binary and low-dimensional, SVM training is a convex program with a unique global optimum, and SVMs generalize reliably from modest training sets \cite{jin2021machine,MURATKAR2024155222,8761796,zhu2025blind}.


Early AmBC receivers relied on energy detection and differential modulation, in which non-coherent schemes requiring minimal receiver complexity recover the tag bit by comparing the average received power to an adaptive threshold \cite{liu2013ambient,8368232}. Such designs are attractive for ultra-low-power operation, but their performance is limited by the dominant direct-link contribution and by sensitivity to noise and fading, particularly at low signal-to-noise ratio (SNR) \cite{Wang2016AmbientBackscatter}. To improve reliability, coherent detection strategies that exploit structured channel knowledge were developed in \cite{zheng2020ambient}, while a range of works examined modulation and signal design choices to harden the link against fading \cite{yang2018modulation}. A widely used baseline, and the one most relevant here, is the MMSE detector \cite{8274950,8761796}, which estimates the ambient symbol under each tag hypothesis and then forms an energy-based decision statistic. The MMSE framework is statistically principled and serves as a strong reference, but its coherent operation presumes pilot-aided or estimable channels assumptions that are violated precisely in the obstructed regimes targeted in this paper, motivating the addition of both environmental control and a learning-based decision rule.

An RIS reshapes propagation by introducing programmable, low-cost reflecting elements that introduce controllable phase shifts \cite{wu2019intelligent,DiRenzo2020RIS,basar2020reconfigurable,elmossallamy2020ris}. In the backscatter context, RIS has been investigated as a means to strengthen the effective backscatter link budget and to improve the direct link of the tag \cite{10409611,10200237,10269022,academia22}. By aligning the phases of the cascaded source-RIS-tag path, an RIS can deliver coherent gain that compensates for weak or shadowed direct links, thereby extending coverage into NLoS or cluttered environments. Existing RIS-aided AmBC studies, however, predominantly retain conventional non-coherent or energy-based detectors and consider single-tag or limited-multiplicity settings, and they generally do not exploit the richer feature separability that an optimized RIS confers on a learning-based receiver. The present work departs from this trend by co-designing the RIS configuration with a residual-feature SVM and by explicitly quantifying how the RIS-induced SINR improvement translates into reduced classification error.

Spatial processing at the reader is a second lever for combating direct-link interference and weak backscatter. Receive beamforming with a uniform linear array (ULA) suppresses interference from undesired directions while reinforcing the desired signal, and has been applied in AmBC to improve detection reliability \cite{11012217}. Classical combiners such as maximum-ratio combining (MRC) maximize the array gain in the look direction, while interference-rejecting beamformers trade gain for nulling \cite{goldsmith2005wireless}. Reader-side beamforming is particularly appealing for AmBC because complexity and energy are concentrated at the reader rather than the passive tag. A multiple-tag AmBC system combining beamforming with ML-based symbol detection was proposed in \cite{11012217}, demonstrating that spatial filtering and learning are complementary. The architecture in this paper uses two receive beamformers in separate stages, one steered toward the RF source for symbol reconstruction and one toward the tag for backscatter collection, and integrates them with RIS-assisted link and residual-feature classification.

ML has been increasingly adopted for AmBC detection because the multiplicative, doubly faded backscatter channel and the nonlinear mixing of direct and reflected components make purely model-based detectors brittle when channel knowledge is incomplete \cite{8761796,jin2021machine}. Neural-network detectors were applied to AmBC under nonlinear impairments in \cite{jin2021machine}, and the comparative study in \cite{MURATKAR2024155222} showed that shallow classifiers such as SVMs and $k$-nearest neighbors can effectively mitigate residual interference in complex propagation settings. Among these, the SVM is adopted in this work for three reasons that align with the structure of the tag-detection problem. First, the decision is intrinsically binary, and the discriminative information is contained in a compact two-dimensional feature space (the pair of residual energies), so the high-capacity function approximation of deep networks is unnecessary, whereas the margin-maximizing boundary of an SVM matches the geometry directly. Second, SVM training is a convex optimization with a unique global optimum, so the detector is reproducible across runs and does not depend on initialization or stopping criteria, which makes BER comparisons clean to interpret \cite{cortes1995support}. Third, SVMs remain effective with the modest training sets available in AmBC and generalize reliably in low-dimensional feature spaces \cite{8761796,MURATKAR2024155222}. Prior ML-based AmBC detectors, however, typically assume an uncontrolled propagation environment and train per operating point, whereas we instead co-design the RIS to enhance separability and train across multiple SINR levels for robustness. 
Finally, several works highlight that AmBC performance is acutely sensitive to channel geometry and line-of-sight conditions. When the direct path to the reader is blocked, the reader encounters severe difficulty decoding the backscattered signal \cite{8368232}, and the weak, doubly faded backscatter component is easily lost in deep fades \cite{Wang2016AmbientBackscatter,zheng2020ambient}. This sensitivity is the central practical obstacle to deploying AmBC in indoor, industrial, and body-area scenarios, where obstruction and shadowing are the norm rather than the exception. 


A review of the literature above reveals that the combined exploitation of RIS-aided propagation control and ML-based detection in AmBC remains underexplored, and that obstructed-propagation operation has received little dedicated attention. RIS-aided AmBC frameworks tend to retain traditional coherent or non-coherent detectors, overlooking the potential of ML to exploit improved channel conditions enabled by an RIS. Conversely, ML-based AmBC detectors generally assume a static or uncontrolled propagation environment, thereby neglecting the opportunity to co-design the RIS to enhance feature separability for classification. Furthermore, although residual energy metrics arising from hypothesis-based equalization have been used as discriminative features in other contexts \cite{Wang2016AmbientBackscatter}, their integration into an RIS-aided AmBC receiver, and the explicit link between the resulting feature separability and SINR, have not been established. This paper addresses these gaps and, unlike prior studies that treat RIS or ML in isolation, assume favorable direct links, or retain conventional detectors, we consider an RIS-aided, beamforming-assisted, residual-feature SVM detector for AmBC under obstructed and deep-fading propagation, and we explicitly account for the coupling between physical-layer SINR optimization and the separability of the learned decision. 
The main contributions of this work are summarized below:
\begin{itemize}
\item We consider an RIS-aided AmBC system in which the direct source-to-tag and source-to-reader links are obstructed or deeply faded, and the RIS phase profile is aligned to strengthen the cascaded source-RIS-tag link. 
\item  We reformulate the equalization stage so that, for each candidate tag state, the receiver reconstructs the ambient symbol via MMSE and computes the corresponding residual energy. The pair of residual energies forms a compact, physically interpretable feature vector for classification, and we make explicit how these features are produced and why they separate the tag states.
\item  We clarify the relationship between physical-layer optimization and learning by showing that RIS phase alignment and receive beamforming raise the SINR, which in turn increases the separation between the two residual-feature clusters and thereby lowers the BER.
\item 
Through extensive simulation, we show consistent BER gains over RIS-energy, SVM, and SVM-beamforming baselines, quantify the relative influence of RIS size and antenna count, and demonstrate graceful degradation under obstruction and across SINR.
\end{itemize}

The remainder of the paper is organized as follows.  Section~\ref{sec:sysmodel} presents the system, channel, and signal models, including the RIS phase design, receive beamforming, and the SINR-based spectral-efficiency upper bound. Section~\ref{sec:mmse} develops the hypothesis-based MMSE equalizer and residual-feature generation. Section~\ref{sec:svm} describes the SVM detector, the SINR-to-separability coupling, and complexity. Section~\ref{sec:results} reports simulation results, and Section~\ref{sec:conclusion} concludes the paper.


\section{System Model}\label{sec:sysmodel}
\begin{figure}[!t]
    \centering
    \includegraphics[width=0.99\columnwidth]{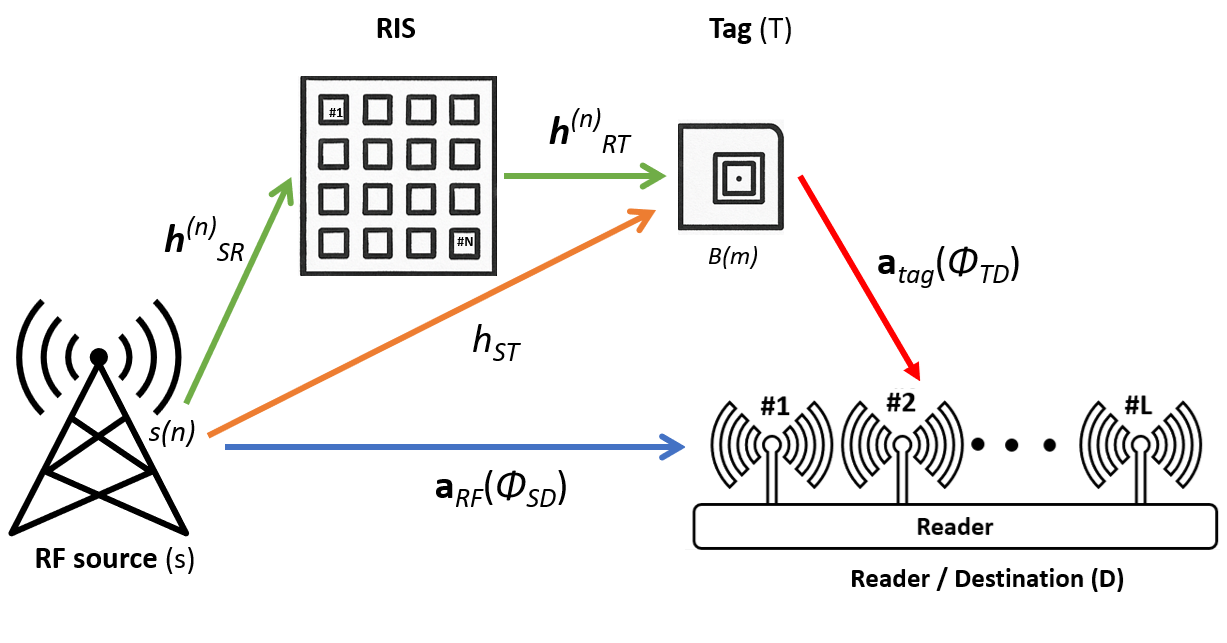}
    \caption{RIS-aided AmBC system with reader-side beamforming and residual-feature SVM detection.}
    \label{fig:system_model}
\end{figure}

We consider the RIS-aided AmBC system illustrated in Fig.~\ref{fig:system_model}, comprising an RF source $S$, a passive backscatter tag $T$, a reader $D$ equipped with an $L$-element ULA, and an RIS with $N_R$ reflecting elements. The notation used throughout is summarized in Table~\ref{tab:notation}. The RF source transmits a quadrature phase-shift keying (QPSK) ambient signal $s_m(n)$, which illuminates both the tag and the RIS. Here $n$ is the RF-symbol time index and $m$ is the tag-symbol index. During each tag-symbol duration the source emits $N$ RF symbols, so that $n=1,\dots,N$ and $m=1,\dots,M$. The passive tag modulates its binary information symbol $B(m)\in\{-1,+1\}$ onto the incident carrier by varying its antenna impedance, thereby altering its reflection coefficient and embedding the tag data into the backscattered waveform. The RIS adjusts the phase of each of its $N_R$ reflecting elements to strengthen the cascaded link of the tag, improving the amplitude and reliability of the backscatter before it reaches the reader. The reader applies receive beamforming to steer its reception toward the RF source and the tag, enhancing the desired components while suppressing interference from undesired directions.

\begingroup
\begin{table}[!t]
\centering
\caption{Summary of Notation}
\label{tab:notation}
\renewcommand{\arraystretch}{1.2}
\begin{tabular}{@{}c l@{}}
\toprule
Symbol & Description \\
\midrule
$S,T,D$ & RF source, tag, reader \\
$L,\,N_R$ & Number of reader antennas, RIS elements \\
$N,\,M$ & RF symbols per tag symbol, number of tag symbols \\
$s_m(n)$ & QPSK ambient symbol ($n$ within tag symbol $m$) \\
$B(m)$ & BPSK tag symbol, $B(m)\in\{-1,+1\}$ \\
$\zeta,\,\eta$ & Tag reflection coefficient, RIS efficiency \\
$h_{ST}$ & Direct source-to-tag channel \\
$h_{SD}$ & Direct source-to-reader channel \\
$h_{TD}$ & Tag-to-reader (backscatter) channel \\
$h_{SR}^{(q)}$ & Source-to-RIS-element-$q$ channel \\
$h_{RT}^{(q)}$ & RIS-element-$q$-to-tag channel \\
$\theta^{(q)}$ & Phase shift of RIS element $q$ \\
$\boldsymbol{\Theta}$ & RIS reflection matrix $\diag(e^{j\theta_1},\dots,e^{j\theta_{N_R}})$ \\
$\phi_{SD},\,\phi_{TD}$ & Angles of arrival of source and tag \\
$\mathbf{a}_{RF},\,\mathbf{a}_{tag}$ & Source and tag array-response vectors \\
$\mathbf{w}_{RF},\,\mathbf{w}_{tag}$ & Receive beamformers (source, tag) \\
$g_{ST}$ & Effective source-to-tag channel (RIS $+$ direct) \\
$\gamma$ & Instantaneous SINR at the reader \\
$\epsilon(m)_{\pm}$ & Residual energies under the two hypotheses \\
$\mathbf{x}_m$ & Feature vector $[\epsilon(m)_-,\,\epsilon(m)_+]^T$ \\
\bottomrule
\end{tabular}
\end{table}
\endgroup

Let $h_{SD}\in\mathbb{C}$ denote the direct source-to-reader channel, $h_{TD}\in\mathbb{C}$ the tag-to-reader backscatter channel, and $h_{ST}\in\mathbb{C}$ the direct source-to-tag channel. The source-to-RIS and RIS-to-tag channels are represented as vectors $\mathbf{h}_{SR}=[h_{SR}^{(1)},\dots,h_{SR}^{(N_R)}]^T\in\mathbb{C}^{N_R\times1}$ and $\mathbf{h}_{RT}=[h_{RT}^{(1)},\dots,h_{RT}^{(N_R)}]^T\in\mathbb{C}^{N_R\times1}$, respectively. All links experience independent and identically distributed (i.i.d.) Rayleigh block fading, so that each coefficient is a circularly symmetric complex Gaussian random variable and the instantaneous power gains are exponentially distributed \cite{liu2013ambient,zheng2020ambient}.
Consistent with the exponential power model, the instantaneous received SINR $\gamma$ has the probability density function
\begin{equation}
f_{\gamma}(\gamma)=\frac{1}{\bar{\gamma}}\exp\!\lb(-\frac{\gamma}{\bar{\gamma}}\rb),\qquad \gamma\geq0,
\label{eq:pdf}
\end{equation}
where $\bar{\gamma}$ denotes the average SINR.

\subsection{RIS Phase Configuration}
The RIS acts as a controllable reflector that introduces an additional source-RIS-tag route, enabling phase-aligned constructive combining at the tag. For  $q$-th element, the signal traverses $h_{SR}^{(q)}$, acquires the reflection phase $\theta^{(q)}$, and traverses $h_{RT}^{(q)}$, so the per-element cascade gain is $h_{SR}^{(q)}e^{j\theta^{(q)}}h_{RT}^{(q)}$. To align every element with the direct source-to-tag path, the optimal phase of element $q$ is set to

\begin{equation}
\theta^{(q)}=\angle h_{ST}-\lb(\angle h_{SR}^{(q)}+\angle h_{RT}^{(q)}\rb),
\label{eq:phase}
\end{equation}
so that each cascade term has phase $\angle h_{ST}$ and adds coherently with the direct component. The RIS reflection response is the diagonal matrix
\begin{equation}
\boldsymbol{\Theta}^{\star}=\diag\!\lb(e^{j\theta_1^{\star}},e^{j\theta_2^{\star}},\dots,e^{j\theta_{N_R}^{\star}}\rb).
\label{eq:Theta}
\end{equation}
The resulting effective source-to-tag channel, combining the RIS-reflected cascade and the (possibly obstructed) direct path, is
\begin{equation}
g_{ST}=\eta\,\mathbf{h}_{SR}^{H}\boldsymbol{\Theta}^{\star}\mathbf{h}_{RT}+h_{ST},
\label{eq:gST}
\end{equation}
where $\eta\in(0,1]$ is the RIS amplitude efficiency. When the direct channel $h_{ST}$ is deep faded, the coherent cascade $\eta\,\mathbf{h}_{SR}^{H}\boldsymbol{\Theta}^{\star}\mathbf{h}_{RT}$ dominates $g_{ST}$ {\cite{chen2021performance, wafai_ojcoms, wafai_tvt}}.


\subsection{Receive Beamforming}
The reader employs a ULA with inter-element spacing $d$ and operating wavelength $\lambda$. The array's response to a signal arriving from angle $\phi$ is captured by a steering vector. Following the ULA processing in \cite{11012217}, the array-response vectors for the RF source at angle $\phi_{SD}$ and the tag at angle $\phi_{TD}$ are
\begin{align}
\mathbf{a}_{RF}(\phi_{SD})
&=h_{SD}\big[1,e^{-j\frac{2\pi d}{\lambda}\sin\phi_{SD}},
\dots,e^{-j\frac{2\pi(L-1)d}{\lambda}\sin\phi_{SD}}\big]^T,
\label{eq:aRF}\\[-1mm]
\mathbf{a}_{tag}(\phi_{TD})
&=h_{TD}\!\big[1,e^{-j\frac{2\pi d}{\lambda}\sin\phi_{TD}},
\dots,e^{-j\frac{2\pi(L-1)d}{\lambda}\sin\phi_{TD}}\big]^T.
\label{eq:atag}
\end{align}
where, following the convention of the system, the direct-link channel gains $h_{SD}$ and $h_{TD}$ are absorbed into the corresponding array-response vectors. To extract the desired signals, the reader applies MRC beamforming, aligning the receive weight with the incoming direction
\begin{equation}
\mathbf{w}_{RF}=\frac{\mathbf{a}_{RF}^{H}(\phi_{SD})}{\lVert\mathbf{a}_{RF}^{H}(\phi_{SD})\rVert},\qquad
\mathbf{w}_{tag}=\frac{\mathbf{a}_{tag}^{H}(\phi_{TD})}{\lVert\mathbf{a}_{tag}^{H}(\phi_{TD})\rVert}.
\label{eq:weights}
\end{equation}
The vector $\mathbf{w}_{RF}$ steers reception toward the ambient RF source, and $\mathbf{w}_{tag}$ steers toward the backscatter signal from the tag. Because the array can steer only a single beam at any instant, the two beamformers are used in separate stages, with $\mathbf{w}_{tag}$ used to collect the backscatter component during tag detection and $\mathbf{w}_{RF}$ used to reconstruct the ambient symbol during equalization (Section~\ref{sec:mmse}).

\subsection{Received Signal}
After receive beamforming toward the tag, the signal at the reader is
\begin{align}\label{eq:rx}
y(n)&=\sqrt{P}\Big[\mathbf{w}_{tag}\mathbf{a}_{RF}(\phi_{SD}) \nonumber
\\&+B(m)\,\mathbf{w}_{tag}\mathbf{a}_{tag}(\phi_{TD})\,\zeta\,g_{ST}\,\Big]s_m(n)+w(n),
\end{align}
where $g_{ST}$ is the effective source-to-tag channel of \eqref{eq:gST}, $\zeta$ is the tag reflection coefficient, and $w(n)\sim\mathcal{CN}(0,\sigma^{2})$ is additive white Gaussian noise. The first term is the residual direct-link contribution that leaks through the tag beamformer, and the second is the desired backscatter component whose sign carries the tag bit $B(m)$. The tag employs BPSK and performs backscatter by switching its load impedance between two states. For analytical convenience we set the transmit power $P=1$ and, treating the residual direct-link leakage as interference, obtain the instantaneous SINR following \cite{Wang2016AmbientBackscatter} as
\begin{equation}
\gamma=\frac{\big|\mathbf{w}_{tag}\mathbf{a}_{tag}(\phi_{TD})\,\zeta\,g_{ST}\big|^{2}}
{\sigma^{2}+\big|\mathbf{w}_{tag}\mathbf{a}_{RF}(\phi_{SD})\big|^{2}}.
\label{eq:sinr}
\end{equation}
The tag symbol $B(m)$ does not appear in \eqref{eq:sinr} because $|B(m)|^{2}=1$. The SINR in ~\eqref{eq:sinr} makes the physical-layer interactions explicit. The RIS phase alignment increases $|g_{ST}|$ through the coherent cascade in \eqref{eq:gST}, while the tag beamformer simultaneously strengthens the desired signal in the numerator.
We define the SINR-based spectral-efficiency upper bound as
\begin{equation}
R_{\mathrm{ub}}=\log_{2}\!\lb(1+\gamma\rb)\quad[\text{bits/s/Hz}].
\label{eq:rate}
\end{equation}
Since the tag transmits BPSK, $R_{\mathrm{ub}}$ is {not} the tag's achievable coded rate. The Gaussian-input expression \eqref{eq:rate} upper-bounds the mutual information of any fixed-power input, and for the BPSK tag, the constrained rate cannot exceed $\min(1,R_{\mathrm{ub}})$ bit/s/Hz. We report $R_{\mathrm{ub}}$ purely as an SINR-based link-quality indicator that isolates the physical-layer gains from RIS phase alignment and receive beamforming. 

\section{Hypothesis-Based MMSE Equalization and Residual-Feature Generation}\label{sec:mmse}

The tag bit $B(m)$ flips the sign of the backscatter term in \eqref{eq:rx}, yielding two candidate effective channels at the reader. Detection therefore proceeds by testing both hypotheses, $B(m)=-1$ and $B(m)=+1$, reconstructing the ambient symbol under each, and measuring how well each reconstruction explains the observation. This section formalizes that procedure and shows how it produces the residual features used by the classifier.

\subsection{Per-Hypothesis Effective Channels}
For symbol reconstruction, the reader uses the source beamformer $\mathbf{w}_{RF}$, since the MMSE stage recovers the ambient waveform $s_m(n)$. Under the two hypotheses, the equivalent channels are
\begin{align}
\mathbf{H}_{-1}^{(m)}&=\mathbf{w}_{RF}\mathbf{a}_{RF}(\phi_{SD})
-\zeta\,\mathbf{w}_{RF}\mathbf{a}_{tag}(\phi_{TD})\,g_{ST},\label{eq:Hminus}\\
\mathbf{H}_{+1}^{(m)}&=\mathbf{w}_{RF}\mathbf{a}_{RF}(\phi_{SD})
+\zeta\,\mathbf{w}_{RF}\mathbf{a}_{tag}(\phi_{TD})\,g_{ST},\label{eq:Hplus}
\end{align}
where $g_{ST}$ is the effective source-to-tag channel of \eqref{eq:gST}. The two channels differ only in the sign of the backscatter contribution, so the correct hypothesis is the one whose reconstructed signal best matches the received samples.

\subsection{MMSE Symbol Reconstruction}
Under each hypothesis, an MMSE equalizer recovers the ambient symbol $s_m(n)$, producing the estimates $\hat{s}_m(n)_{-}$ and $\hat{s}_m(n)_{+}$ for $B(m)=-1$ and $B(m)=+1$, respectively \cite{8274950},
\begin{align}
\hat{s}_m(n)_{-}&=\arg\min_{s_m(n)\in\mathcal{A}_s}\Bigg|\frac{\big(\mathbf{H}_{-1}^{(m)}\big)^{H}}{\lVert\mathbf{H}_{-1}^{(m)}\rVert^{2}}\,y_m(n)-s_m(n)\Bigg|^{2},\label{eq:mmseminus}\\
\hat{s}_m(n)_{+}&=\arg\min_{s_m(n)\in\mathcal{A}_s}\Bigg|\frac{\big(\mathbf{H}_{+1}^{(m)}\big)^{H}}{\lVert\mathbf{H}_{+1}^{(m)}\rVert^{2}}\,y_m(n)-s_m(n)\Bigg|^{2},\label{eq:mmseplus}
\end{align}
where $\mathcal{A}_s$ is the QPSK alphabet and $y_m(n)$ is the received sample at RF-index $n$ within tag symbol $m$. Each estimate is the alphabet symbol closest to the MMSE-equalized observation under the assumed channel.

\subsection{Residual-Energy Features}
To generate discriminative features, the reader reconstructs the received signal under each hypothesis and measures the residual reconstruction energy over the $N$ RF samples of the tag symbol. The residuals under the two hypotheses are
\begin{align}
\epsilon(m)_{-}&=\frac{1}{N}\sum_{n=1}^{N}\Big|y(n)-\mathbf{H}_{-1}^{(m)}\hat{s}_m(n)_{-}\Big|^{2},\label{eq:resminus}\\
\epsilon(m)_{+}&=\frac{1}{N}\sum_{n=1}^{N}\Big|y(n)-\mathbf{H}_{+1}^{(m)}\hat{s}_m(n)_{+}\Big|^{2}.\label{eq:resplus}
\end{align}
{When the assumed hypothesis matches the transmitted bit, the reconstructed signal $\mathbf{H}^{(m)}\hat{s}_m(n)$ closely tracks $y(n)$ and the residual energy is small. When the hypothesis is wrong, the sign mismatch in the backscatter term inflates the residual. The true tag state is therefore associated with the smaller residual, and the \emph{pair} $\big(\epsilon(m)_-,\epsilon(m)_+\big)$ encodes which hypothesis is more consistent with the data. Rather than thresholding this pair directly, which is fragile under fading and residual interference, we treat it as a feature vector.
\begin{equation}
\mathbf{x}_m=\lb[\epsilon(m)_-,\ \epsilon(m)_+\rb]^{T}\in\mathbb{R}^{2},
\label{eq:feature}
\end{equation}
and learn the decision boundary that best separates the two tag states. The construction in \eqref{eq:Hminus}--\eqref{eq:feature} is the residual-feature generation referred to throughout, and it converts the physical received signal into a compact, two-dimensional representation whose geometry is governed by the SINR in \eqref{eq:sinr}.

\section{SVM-Based Detection}\label{sec:svm}
The reader classifies the tag symbols using the residual features in \eqref{eq:feature}. Each instance is the pair $\mathbf{x}_m=[\epsilon(m)_-,\epsilon(m)_+]^{T}$ with label $B_m\in\{-1,+1\}$, where $m=1,\dots,M$ and $M$ is the number of training instances. An SVM seeks the separating hyperplane
\begin{equation}
\mathbf{w}^{T}\mathbf{x}+b=0,
\label{eq:hyperplane}
\end{equation}
where $\mathbf{w}\in\mathbb{R}^{2}$ is the weight vector and $b\in\mathbb{R}$ is the bias. Maximizing the margin between the two classes improves robustness to the channel variations and residual interference that would otherwise cause misclassification.

\subsection{Primal, Dual, and Decision Function}
For linearly separable features the primal problem is
\begin{equation}
\min_{\mathbf{w},b}\ \tfrac{1}{2}\lVert\mathbf{w}\rVert^{2}\quad
\text{s.t.}\quad B_m\big(\mathbf{w}^{T}\mathbf{x}_m+b\big)\geq1,\ \ m=1,\dots,M.
\label{eq:primal}
\end{equation}
Minimizing $\lVert\mathbf{w}\rVert^{2}$ maximizes the margin, which is directly tied to the classifier's ability to generalize under fading. Introducing Lagrange multipliers $\alpha_m\geq0$, the dual problem is
\begin{equation}
\max_{\boldsymbol{\alpha}}\ \sum_{m=1}^{M}\alpha_m-\tfrac{1}{2}\sum_{m=1}^{M}\sum_{j=1}^{M}\alpha_m\alpha_j B_m B_j\,K(\mathbf{x}_m,\mathbf{x}_j),
\label{eq:dual}
\end{equation}
where $j=1,\dots,M$ indexes the dual variables and $K(\cdot,\cdot)$ is the kernel function. The decision function for a test feature $\mathbf{x}$ is
\begin{equation}
f(\mathbf{x})=\sgn\!\lb(\sum_{m=1}^{M}\alpha_m B_m\,K(\mathbf{x}_m,\mathbf{x})+b\rb),
\label{eq:decision}
\end{equation}
where $\sgn(\cdot)$ returns the predicted symbol class, and only the support vectors (those with $\alpha_m>0$) contribute, which keeps inference efficient.

\subsection{Gaussian Kernel}
To accommodate non-linearly separable features, we adopt the Gaussian radial-basis-function (RBF) kernel \cite{scholkopf2002learning}
\begin{equation}
K(\mathbf{u},\mathbf{v})=\exp\!\lb(-\frac{\lVert\mathbf{u}-\mathbf{v}\rVert^{2}}{2\ell^{2}}\rb),
\label{eq:rbf}
\end{equation}
where $\ell$ is the kernel length-scale controlling the smoothness of the boundary. The RBF kernel captures the nonlinear variations induced by multipath fading \cite{jin2021machine}, and tuning $\ell$ balances sensitivity to local structure against generalization, providing reliable detection across fading severities.

\subsection{Coupling Between SINR Optimization and Feature Separability}}
A central observation of this work is that the physical-layer optimization and the learning stage are not independent but are coupled through the geometry of the residual features. Consider the residuals in \eqref{eq:resminus}--\eqref{eq:resplus}. When the SINR $\gamma$ in \eqref{eq:sinr} is high because RIS phase alignment has increased $|g_{ST}|$ and the tag beamformer has both reinforced the backscatter component and suppressed the direct-link leakage, the residual under the correct hypothesis is driven close to the noise floor, while the residual under the incorrect hypothesis is inflated by the now-strong, sign-mismatched backscatter term. The two feature clusters $\{\mathbf{x}_m:B_m=-1\}$ and $\{\mathbf{x}_m:B_m=+1\}$ therefore move apart, and the SVM's margin grows. Conversely, under heavy obstruction and low SINR, the backscatter term is weak, the two residuals are similar regardless of the true bit, and the clusters overlap, so no classifier can separate them well.
This yields a clear chain of causation that the architecture is designed to exploit.
\begin{enumerate}
\item RIS phase alignment \eqref{eq:phase} and receive beamforming \eqref{eq:weights} operate at the physical layer to raise the SINR $\gamma$ in \eqref{eq:sinr}.
\item the MMSE hypothesis stage \eqref{eq:Hminus}--\eqref{eq:resplus} converts the SINR improvement into a larger separation between the residual-feature clusters.
\item the SVM \eqref{eq:primal}--\eqref{eq:decision} exploits this cleaner feature space to achieve a larger margin and hence lower BER.
\end{enumerate}
Higher SINR values generally imply better feature separability, which is why the proposed detector benefits from the RIS and beamforming rather than treating them as separate, decoupled subsystems.

\subsection{Computational Complexity}
The detector's complexity is dominated by RBF-SVM training, which solves the quadratic program \eqref{eq:dual} over $M$ training samples and operates on an $M\times M$ kernel matrix, giving $O(M^{3})$ complexity, where the cubic term arises because the multipliers $\alpha_m$ are updated iteratively using the full kernel matrix. Inference scales as $O(n_{SV}\,p)$, where $n_{SV}$ is the number of support vectors and $p$ is the feature dimension. Here $p=2$, so inference is lightweight. Training is performed offline, which keeps real-time detection at the reader computationally efficient. The MMSE and residual-feature stages contribute $O(N)$ operations per tag symbol for reconstruction and energy accumulation.

\section{Results and Discussion}\label{sec:results}
Simulations were conducted in MATLAB to evaluate the proposed RIS-beamforming-SVM detector. Unless otherwise stated, the number of tag symbols was $M=10{,}000$, each composed of $N=64$ RF source symbols. The tag reflection coefficient and RIS efficiency were $\zeta=1$ and $\eta=0.5$, the RF source and tag were located at $\phi_{SD}=-45^{\circ}$ and $\phi_{TD}=60^{\circ}$, the reader used $L=10$ antennas with normalized spacing $d/\lambda=1$, and the RIS comprised $N_R=64$ elements with phases set by \eqref{eq:phase}. For SVM classification a $50\%/50\%$ training-testing split was used. 

\begin{figure}[!t]
    \centering
    \includegraphics[width=0.95\columnwidth]{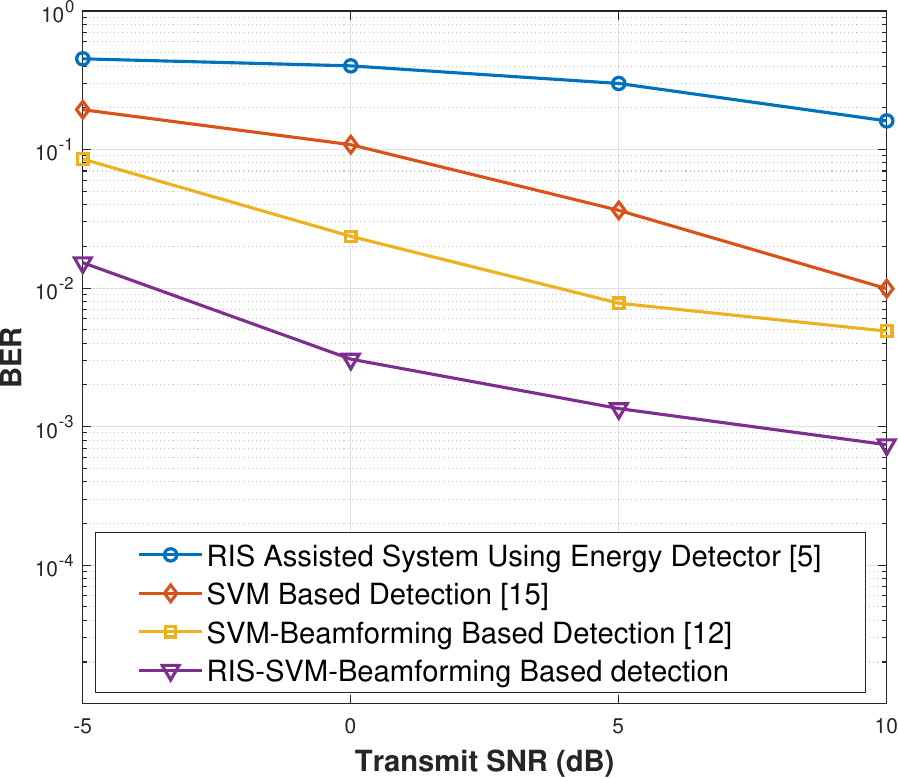}
    \caption{BER comparison of RIS-aided and baseline AmBC architectures under different detection schemes.}
    \label{fig:ber_comparison}
\end{figure}
Fig.~\ref{fig:ber_comparison} compares the proposed hybrid RIS-SVM detector against the RIS-assisted energy detector \cite{10200237,10269022}, the SVM detector \cite{8761796}, and the SVM-beamforming detector \cite{11012217}. The proposed scheme achieves the lowest BER across all SINR values, confirming that integrating RIS-assisted link, receive beamforming, MMSE reconstruction, and SVM classification produces a synergistic improvement. The hybrid design combines channel enhancement through the RIS, coherent symbol reconstruction via MMSE, and feature-based discrimination via the SVM, yielding superior reliability, particularly in the low-SINR regime that characterizes truly passive AmBC.

\begin{figure}[!t]
    \centering
    \includegraphics[width=0.95\columnwidth]{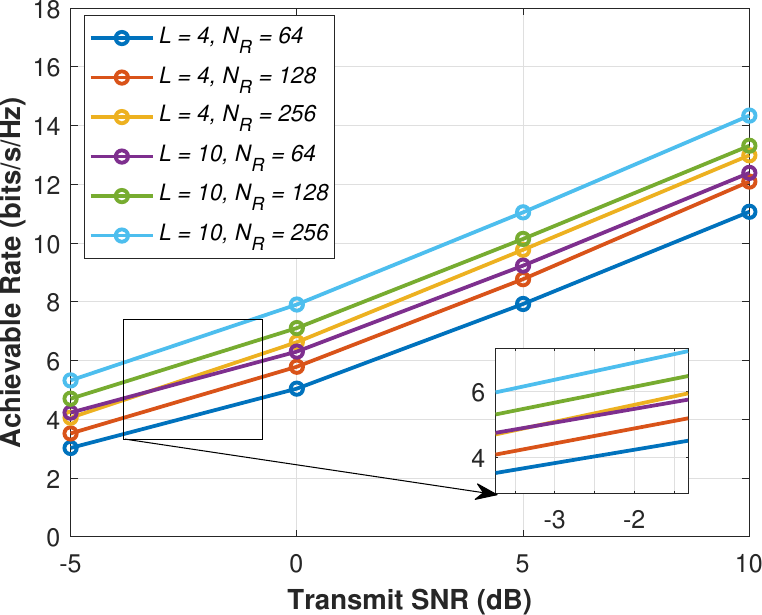}
    \caption{{SINR-based upper bound rate} $R_{\mathrm{ub}}$ versus SNR for the RIS-beamforming-SVM scheme.}
    \label{fig:achievable_rate}
\end{figure}
Fig.~\ref{fig:achievable_rate} reports the SINR-based upper bound $R_{\mathrm{ub}}$ of \eqref{eq:rate} as a function of transmit SNR. At low SNR ($-5$ to $0$~dB), the configuration $L=10$, $N_R=64$ performs comparably to the smaller-array, larger-surface configuration $L=4$, $N_R=256$, indicating that in the noise-limited regime a larger RIS can compensate for fewer reader antennas. As SNR increases, the larger-RIS configuration begins to dominate, demonstrating that in the moderate-to-high SNR regime $R_{\mathrm{ub}}$ is governed more strongly by the RIS size than by the antenna count, since the RIS contributes more to channel enhancement through coherent reflection and added spatial diversity, whereas antenna expansion alone yields diminishing returns. At high SNR, $R_{\mathrm{ub}}$ grows almost linearly for all configurations, with larger RIS deployments yielding the most pronounced gains. {We note that $R_{\mathrm{ub}}$ is an SINR-driven ceiling and not the BPSK tag's coded rate as emphasized in Section~\ref{sec:sysmodel}.}

\begin{figure}[!t]
    \centering
    \includegraphics[width=0.95\columnwidth]{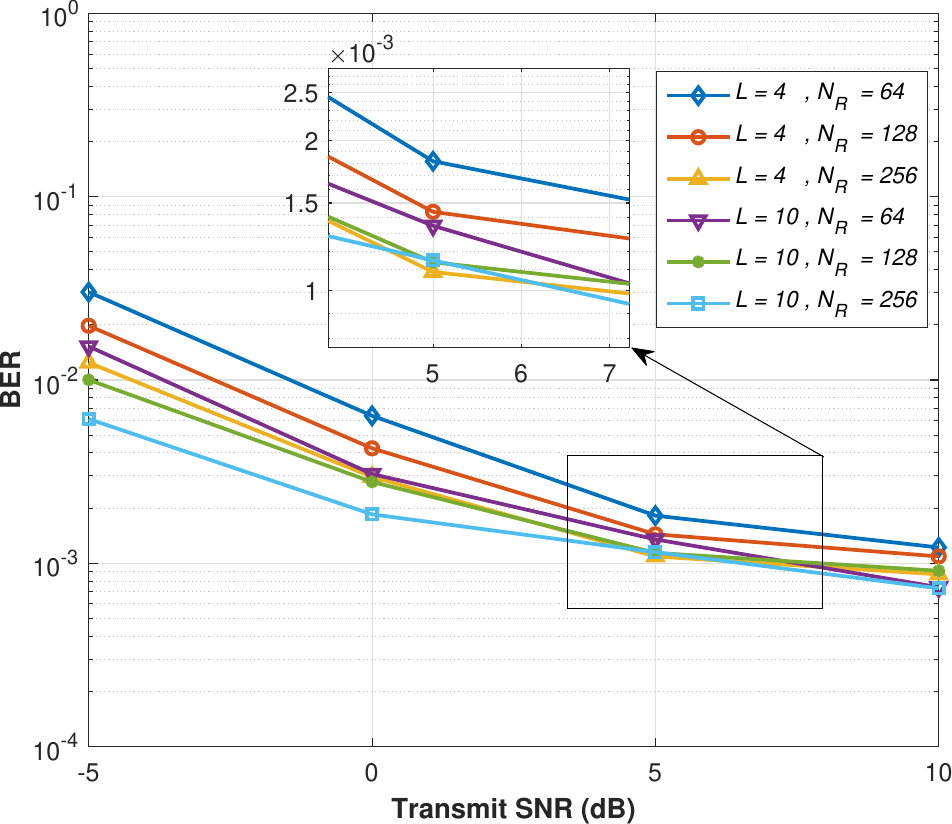}
    \caption{BER of the RIS-beamforming-SVM detector for varying numbers of reader antennas and RIS elements.}
    \label{fig:antenna_IRS_tradeoff}
\end{figure}
Fig.~\ref{fig:antenna_IRS_tradeoff} analyzes the combined impact of the antenna count $L$ and the RIS size $N_R$. Increasing either parameter improves BER through enhanced spatial diversity and stronger effective reflections. A crossover is observed near $\gamma=5$~dB, where different configurations intersect. This is consistent with Fig.~\ref{fig:achievable_rate} and highlights that the RIS contributes most in the low-to-moderate SINR range typical of passive AmBC, where it substantially improves received signal strength and detection fidelity.

\begin{figure}[!t]
    \centering
    \includegraphics[width=0.95\columnwidth]{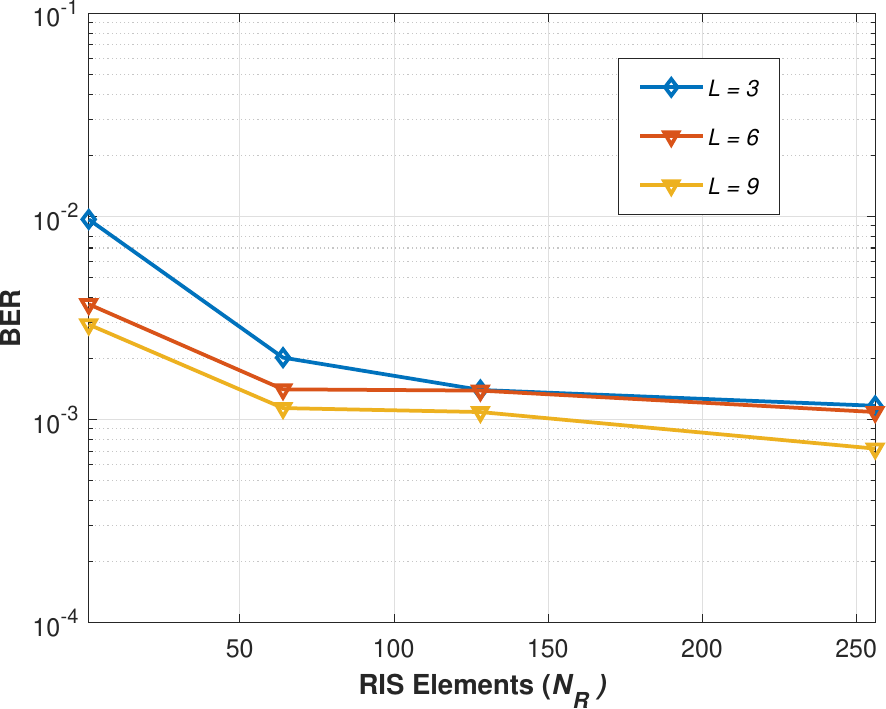}
    \caption{BER at a transmit SNR of $5$~dB for varying numbers of antennas and RIS elements.}
    \label{fig:antenna_IRS_tradeoff1}
\end{figure}
Fig.~\ref{fig:antenna_IRS_tradeoff1} evaluates BER at a fixed transmit SNR of $5$~dB. The BER decreases consistently with increasing RIS size and antenna count, though the improvement diminishes beyond $N_R=128$, with the best result at $N_R=256$. This saturation implies that practical systems can achieve near-optimal performance with a moderate RIS, reducing hardware complexity and cost.

\begin{figure}[!t]
    \centering
    \includegraphics[width=0.95\columnwidth]{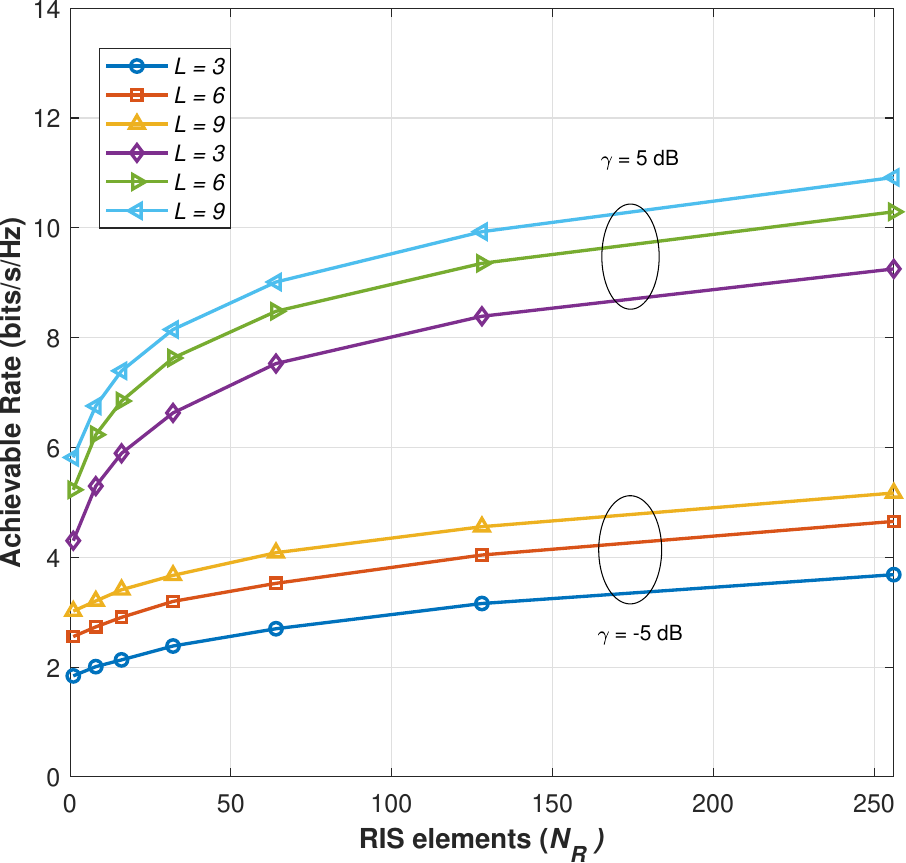}
    \caption{{Achievable rate} versus $N_R$ for different antenna counts at $\gamma=5$~dB and $\gamma=-5$~dB.}
    \label{fig:rate_ris_5}
\end{figure}


Fig.~\ref{fig:rate_ris_5} reports $R_{\mathrm{ub}}$ as a function of $N_R$ for $L\in\{3,6,9\}$ at $\gamma=5$~dB and $\gamma=-5$~dB. The bound increases monotonically with both RIS size and antenna count, since larger surfaces enable finer reflection control and more antennas add spatial diversity. The gain becomes marginal beyond $N_R\approx200$, indicating a diminishing-return region and underscoring the importance of balancing RIS size and array design for cost-effective spectral efficiency.


\section{Conclusion}\label{sec:conclusion}

This paper presented an RIS-aided AmBC framework that combines reader-side beamforming with residual-feature SVM detection to improve reliability under obstructed and deep-fading conditions. By strengthening the cascaded source-RIS-tag path, the proposed method achieves lower BER than the considered baseline schemes. The results show that RIS phase alignment and beamforming improve SINR and enhance the separability of SVM features, thereby improving tag detection reliability. The spectral-efficiency results are interpreted as SINR-based upper bounds on link quality, consistent with the BPSK tag model. Future work will extend the framework to simultaneous multi-tag operation.

\bibliographystyle{IEEEtran}
\bibliography{references}

\begin{IEEEbiography}[{\includegraphics[width=1in,height=1.25in,clip,keepaspectratio]{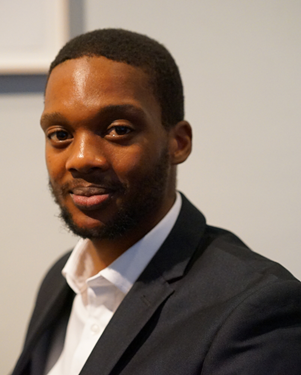}}]{Aobakwe Makgarapa} \vspace{0.01cm}
 (Student Member, IEEE) is currently pursuing the Ph.D. degree with the Wolfson School of Mechanical, Electrical and Manufacturing Engineering, Loughborough University, Loughborough, U.K. He is also a Lecturer at East Midlands Institute of Technology, Loughborough College Group, Loughborough, U.K. His research focuses on wireless communications, ambient backscatter communications, reconfigurable intelligent surfaces, beamforming, and machine-learning-based detection for low-power Internet of Things networks. His broader academic and professional interests also include signal processing, intelligent communication systems, and sustainable engineering applications.
\end{IEEEbiography}

\begin{IEEEbiography}[{\includegraphics[width=1in,height=1.25in,clip,keepaspectratio]{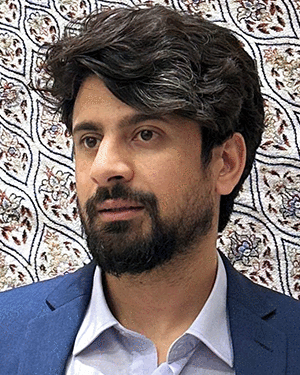}}]{Burhan Wafai}
 \vspace{0.01cm} (Graduate Student Member, IEEE)  
  is currently a Lecturer at East Midlands Institute of Technology, Loughborough College Group, Loughborough, U.K.
 He received his Ph.D. degree from the Department of Electrical Engineering, Indian Institute of Technology (IIT) Jammu, Jammu and Kashmir, India, in 2024. From August 2024 to November 2024, he was a Research Intern with the Department of Computer Science and Information Engineering, National Chung Cheng University, Chiayi, Taiwan. He was an Early Postdoctoral Fellow in the Department of Electrical Engineering at IIT Jammu, India, from December 2024 to February 2025 and a Research Associate with the Wolfson School of Mechanical, Electrical and Manufacturing Engineering, Loughborough University, U.K., from February 2025 to March 2026.  His research interests include wireless communication, physical layer security, integrated sensing and communication, reconfigurable intelligent surfaces, cognitive radio, and resource allocation.
\end{IEEEbiography}

\begin{IEEEbiography}[{\includegraphics[width=1in,height=1.25in,clip,keepaspectratio]{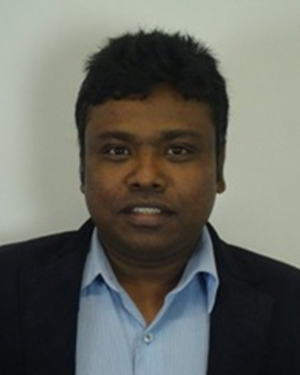}}]{Sangarapillai Lambotharan}
\vspace{0.01cm} (Senior Member, IEEE) is a Professor of Signal Processing and Communications and Director of the Institute for Digital Technologies at Loughborough University London. He received his PhD in Signal Processing from Imperial College London, U.K., in 1997. In 1996, he was a Visiting Scientist at the Engineering and Theory Centre, Cornell University, USA. Until 1999, he was a Post-Doctoral Research Associate at Imperial College London. From 1999 to 2002, he worked with the Motorola Applied Research Group, U.K., where he investigated various projects, including physical link layer modelling and performance characterization of 2.5G and 3G networks. He served as a Lecturer at King's College London and a Senior Lecturer at Cardiff University from 2002 to 2007. His current research interests include 6G networks, OTFS, MIMO, signal processing, machine learning, and network security. He has authored 300 journal articles and conference papers, which have been cited more than 7800 times. He is a Fellow of the IET and a Senior Member of the IEEE. He currently serves as a Senior Area Editor for IEEE Transactions on Signal Processing and served as an Associate Editor for IEEE Transactions on Communications.
\end{IEEEbiography}

\begin{IEEEbiography}[{\includegraphics[width=1in,height=1.25in,clip,keepaspectratio]{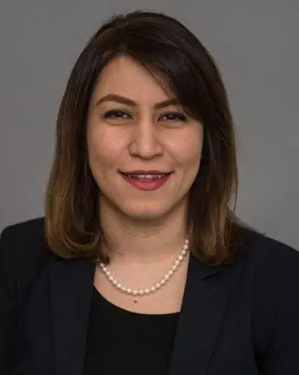}}]{Mahsa Derakhshani}
\vspace{0.01cm} Mahsa Derakhshani (Senior Member, IEEE) received the Ph.D. degree in electrical engineering degree from McGill University, Montr{\'e}al, Canada, in 2013. She was an Honorary NSERC Postdoctoral Fellow with the Department of Electrical and Electronic Engineering, Imperial College London, from 2015 to 2016, a Post-Doctoral Research Fellow with the Department of Electrical and Computer Engineering, University of Toronto, Toronto, Canada, and a Research Assistant with the Department of Electrical and Computer Engineering, McGill University, from 2013 to 2015. She is currently a Reader (Associate Professor) in digital communications with the Wolfson School of Mechanical, Electrical and Manufacturing Engineering, Loughborough University, U.K. Her research interests include machine learning and optimization for wireless communications, ultra-reliable low latency communications, edge computing, and Ambient IoT.
She received several awards and fellowships, including Royal Academy of Engineering/The Leverhulme Trust Research Fellowship (2020--2021), the Natural Sciences and Engineering Research Council of Canada (NSERC) Post-Doctoral Fellow- ships (2015--2017), 
the Fonds de Recherche du Qu{\'e}bec--Nature et Technologies (FRQNT) Post-Doctoral Fellowship (2013--2015), the John Bonsall Porter Prize (2009--2010), and the McGill Engineering Doctoral Award (2008--2011).
She serves as an Editor for IEEE WIRELESS COMMUNICATIONS LETTERS, IEEE Internet of Things Magazine, and IET Signal Processing journal.
\end{IEEEbiography}

\end{document}